\documentclass[]{aa}
\usepackage{graphicx}
\usepackage{amssymb}
\begin{document}
   \title{New seismological results on the G0\,IV $\eta$ Bootis\thanks{Based
on observations obtained at the 1.2-m
Swiss Euler telescope at La Silla (Chile) and at the 1.93-m telescope at the
Haute-Provence Observatory (France)}}

   \author{F. Carrier\inst{1} \and P. Eggenberger\inst{1} \and F. Bouchy\inst{1,2}}
   \institute{Observatoire de Gen\`eve, 51 chemin de Maillettes,
             CH-1290 Sauverny, Switzerland 
	     \and Laboratoire d'Astrophysique de Marseille, Traverse du Siphon, BP 8, 13376 Marseille Cedex 12, France}
       
   \offprints{F. Carrier\\
   \email{fabien.carrier@obs.unige.ch}}
  
  \date{Received ; accepted }
  \abstract{Several attempts have been made to detect solar-like oscillations in the 
G0 IV star \object{$\eta$~Boo}. We present here new observations on this star
simultaneously conducted with two spectrographs: \textsc{Coralie} 
mounted on the 1.2-m Swiss telescope at the ESO La Silla Observatory (Chile) 
and \textsc{Elodie} based on the {1.93-m} telescope at the Observatoire de Haute 
Provence (France). In total, 1239 spectra were collected over 13 nights.
The power spectrum of the high precision velocity time 
series clearly presents several identifiable peaks between 0.4 and 1.0~mHz showing 
regularity with a large and small separations of $\Delta\nu$\,=\,39.9\,$\mu$Hz and 
$\delta\nu_{02}$\,=\,3.95\,$\mu$Hz respectively. Twenty-two individual 
frequencies have been identified. Detailed models based on these measurements and 
non--asteroseismic observables were computed
using the Geneva evolution code including shellular rotation and atomic diffusion. 
By combining these seismological data with non--asteroseismic observations, we determine
the following global parameters for \object{$\eta$~Boo}:
a mass of $1.57 \pm 0.07$\,$M_{\odot}$, an age $t=2.67 \pm 0.10$\,Gyr and an initial metallicity $(Z/X)_{\mathrm{i}}=0.0391 \pm 0.0070$.
We also show that the mass of \object{$\eta$ Boo} is very sensitive to the choice of the observed metallicity, while
the age of \object{$\eta$ Boo} depends on the input physics used. 
Indeed, a higher metallicity favours a higher mass, while
non--rotating models without overshooting predict a smaller age.
}
  \maketitle
%

\section{Introduction}
The measurements of the frequencies of p-mode
oscillations provide an insight into the internal structure of stars and are nowadays
the most powerful constraint to the theory of stellar evolution.
The five-minute oscillations in the Sun have led to a wealth of information
about the solar interior. These results stimulated various attempts to detect
a similar signal on other solar-like stars by photometric or equivalent
width measurements, with little success due to the extreme weakness of the
expected amplitude. These past years, the stabilized spectrographs developed
for extra-solar planet detection achieved accuracies needed for solar-like 
oscillation detection by means of radial velocity measurements
(Carrier et al. \cite{carrier1}, Bouchy \& Carrier \cite{bouchy1}). 

A primary target for the search for p-mode oscillations is the well-studied bright subgiant G0 
\object{$\eta$ Bootis} (HR5235). Several attempts have been made to detect solar-like oscillations in the 
G0 IV star \object{$\eta$ Boo}. The first result was obtained by Kjeldsen et al. (\cite{kjel1})
with observations conducted with the 2.5-m Nordic Optical Telescope (NOT) on La Palma. 
In contrast to all other detections which were based on velocity measurements
obtained using high-dispersion spectrographs with stable references, they monitored changes
in the equivalent widths (EW) of temperature-sensitive spectral lines. This enabled them to determine
a large separation of 40.3\,$\mu$Hz. Meanwhile, a search for velocity oscillations in \object{$\eta$ Boo}
using the AFOE spectrograph
by Brown et al. (\cite{brown}) has failed to detect a stellar signal.
The analysis of NOT data was refined by Kjeldsen et al. (\cite{kjel2})
using all existing complementary data: new EW measurements obtained with the NOT, new
radial velocities (RV) measured at Lick Observatory and RVs from Brown with the AFOE spectrograph.
They found a large separation of $\Delta\nu$\,=\,40.4\,$\mu$Hz and identified 21 oscillation
frequencies.

In this paper, we report Doppler observations of \object{$\eta$ Boo}tis made with the \textsc{Coralie} and
the \textsc{Elodie} spectrographs in a multi-sites configuration. These new measurements confirm the detection of p-modes
and enable the identification of twenty-two individual mode frequencies, which are compared with those
independently identified by Kjeldsen et al. (\cite{kjel2}). We also present new models of \object{$\eta$ Boo} based on our seismological
constraints. The observations and data
reduction are presented in Sect.~\ref{datareduc}, the acoustic spectrum analysis and the mode identification
in Sect.~\ref{asa}, the calibration and modeling of \object{$\eta$ Boo} in Sect.~\ref{mod}, and the conclusion is given in Sect.~\ref{conc}.
\section{Observations and data reduction}
\label{datareduc}
\begin{figure}
\resizebox{\hsize}{!}{\includegraphics{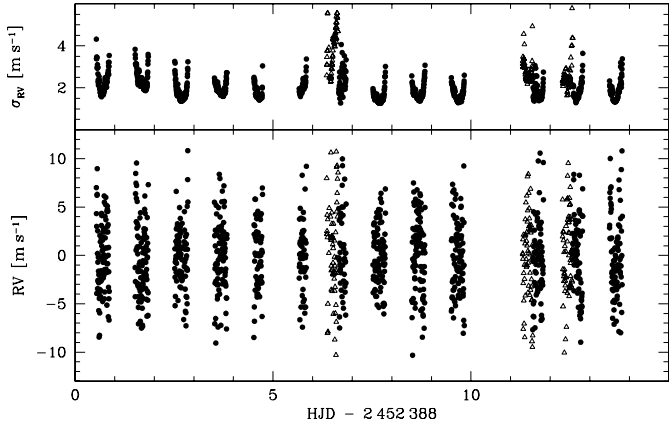}}
\caption{Radial-velocity measurements of \object{$\eta$ Boo}. The dispersion
reaches 3.8\,m\,s$^{-1}$. \textsc{Coralie} and \textsc{Elodie} data are represented with full circles and open
triangles respectively. The upper panel represents the photon noise uncertainties.}
\label{fig:vr}
\end{figure}

\object{$\eta$ Boo} was observed in May 2002 simultaneously with the spectrographs 
\textsc{Coralie} at La Silla Observatory (Chile) and \textsc{Elodie} 
at the Observatoire de Haute 
Provence (France) in order to improve the window function and to make the mode identification easier (see Sect.~\ref{asa}). 

\subsection{\textsc{Coralie} measurements}
\object{$\eta$ Boo} was observed over fourteen nights (April 23 -- May 07 2002) with \textsc{Coralie}, the high-resolution (50\,000)
echelle spectrograph mounted on the 1.2-m Swiss telescope at La Silla, known for the p-mode identification in
the $\alpha$~Cen system (Bouchy \& Carrier \cite{bouchy3}, Carrier \& Bourban \cite{carrier3}).
During the stellar exposures, the spectrum of a
thorium lamp carried by a second fiber is simultaneously recorded in order to
monitor the spectrograph's stability and thus to obtain high-precision
velocity measurements. A description of the spectrograph and the data reduction process is
presented in Carrier et al. (\cite{carrier2}) and Bouchy et al. (\cite{bouchy2}).
Exposure times of 180\,s, thus cycles of 295\,s with a dead-time of 115\,s, allowed us to obtain 1055 spectra,
with a typical signal-to-noise ratio ($S$/$N$) in the range of 115-260 at 550\,nm.
For each night, radial velocities were computed relative to the
highest signal--to--noise ratio optical spectrum obtained in the middle of the
night. The mean for each night is then subtracted.
The radial velocity measurements are shown in Fig.~\ref{fig:vr} and their distribution and 
dispersion are listed in Table~\ref{tab:vr}. The dispersion of these measurements reaches 3.6\,m\,s$^{-1}$.

\subsection{\textsc{Elodie} measurements}
Due to bad weather, only 184 spectra were collected over 3 nights with the high resolution (42\,000) 
spectrograph \textsc{Elodie} 
(Baranne et al. \cite{baranne}) mounted on the 1.93-m telescope at Haute-Provence Observatory (France).
The observations are also achieved in simultaneous-Thorium mode. The wavelength coverage of the spectra
is 3890--6815\,\AA, recorded on 67 orders. The dead-time is about 120\,s as the exposure time varies between
150 and 240\,s depending on the seeing, the extinction and the airmass in order to reach a 
$S$/$N$ in the range 150--300 at 550\,nm. The radial velocity determination method is the same than for
\textsc{Coralie} data. The radial velocity measurements are shown in Fig.~\ref{fig:vr} and their distribution and 
dispersion are listed in Table~\ref{tab:vr}. The dispersion of these measurements reaches 4.3\,m\,s$^{-1}$.

\begin{table}
\caption{Distribution and dispersion of Doppler measurements. Indications for \textsc{Elodie} measurements
are in brackets.}
\begin{center}
\begin{tabular}{clll}
\hline
\hline
Date & No spectra & No hours & $\sigma$ (m\,s$^{-1}$) \\ \hline
2002/04/23 & 96 & 7.83  & 3.69 \\
2002/04/24 & 98 & 8.00  & 3.87 \\
2002/04/25 & 86 & 7.96  & 3.17 \\
2002/04/26 & 93 & 7.54  & 3.60 \\
2002/04/27 & 64 & 5.23  & 3.34 \\
2002/04/28 & 51 & 4.35  & 3.61 \\
2002/04/29&46 (62)&11.81& 4.52 (4.81)  \\
2002/04/30 & 93 & 7.59  & 2.97 \\
2002/05/01 & 99 & 8.02  & 3.86 \\
2002/05/02 & 99 & 7.83  & 3.56 \\
2002/05/03 & -- & --    &   -- \\
2002/05/04&68 (61)&12.61& 3.98 (4.21) \\
2002/05/05&71 (61)&12.26& 3.79 (4.01) \\
2002/05/06 & 91  &7.77  & 4.04 \\ \hline
\label{tab:vr}
\end{tabular}
\end{center}
\end{table}

\section{Power spectrum analysis}
\label{asa}

\begin{figure}
\resizebox{\hsize}{!}{\includegraphics{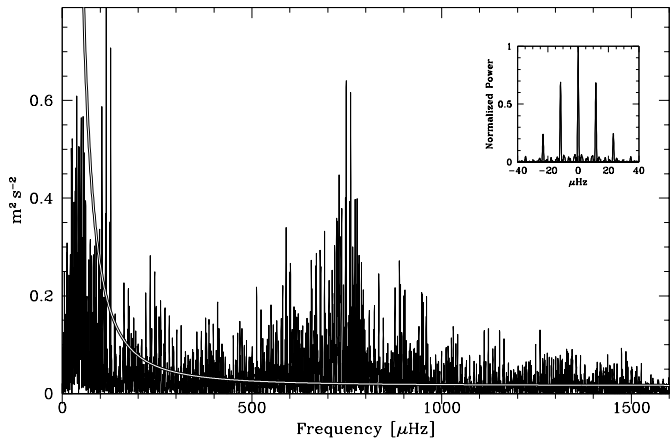}}
\caption{Power spectrum of the radial velocity measurements of \object{$\eta$ Boo}.
The window function is shown in the inset. The noise is represented by the white line.}
\label{fig:tf}
\end{figure}

In order to compute the power spectrum of the velocity time series, we use
the Lomb-Scargle modified algorithm (Lomb \cite{lomb}; Scargle \cite{scargle})
with a weight being assigned to each
point according its uncertainty estimate. The time scale gives a formal resolution
of 0.87\,$\mu$Hz. The resulting periodogram, shown in Fig.~\ref{fig:tf},
exhibits a series of peaks near 0.8\,mHz, exactly where
the solar-like oscillations for this star are expected. 
Typically for such a power spectrum, the noise has two components:
\begin{itemize}
\item At high frequencies it is flat, indicative
of the Poisson statistics of photon noise. The mean white noise level $\sigma_{\mathrm{pow}}$ calculated between 1.2 and 1.6~mHz 
is 0.0179\,m$^2$\,s$^{-2}$, namely 11.9\,cm\,s$^{-1}$ in amplitude. With 1239 measurements, this high frequency noise 
corresponds to $\sigma_{RV}\,=\,\sqrt{N \sigma_{\mathrm{pow}} /4 }\,=\,2.35$~m\,s$^{-1}$. This radial velocity uncertainty,
larger than for others stars with the spectrograph \textsc{Coralie} (see e.g. Eggenberger et al. \cite{eggen1} or 
Bouchy \& Carrier \cite{bouchy3}), can be mainly explained by the high rotational velocity of \object{$\eta$ Boo} ($v \sin i = 12.8$\,km\,s$^{-1}$, determined
from \textsc{Coralie} spectra).
\item Towards the lowest frequencies, the power should scale inversely with frequency squared as expected for instrumental instabilities. 
However, the computation of the radial velocities introduces a high pass filter. Indeed, the radial velocities were computed relative to one
reference for each night and the average radial velocities of the night fixed to zero (see Sect.~\ref{datareduc}). This results in an 
attenuation of the very low frequencies which can be seen on Fig.~\ref{fig:tf}.
\end{itemize}
The power spectrum presents an excess in the range 0.4--1.1~mHz. 
The combined noise has a value decreasing from 0.027 to 0.019\,m$^2$\,s$^{-2}$ (14.5 to 12.2\,cm\,s$^{-1}$) in the above
mentioned interval (see Fig.~\ref{fig:tf}).
The noise has been determined by fitting a function of the type $1/\nu^2$.
Note that the filtering induced by the radial velocities computation does not influence the frequency of the peaks in the 
range 0.4--1.1~mHz, but could slightly change their amplitudes.
 The amplitude of the strongest peaks reaches 79~cm\,s$^{-1}$, corresponding to a signal to noise of 6 (in the amplitude spectrum).
This amplitude is estimated as the height of the peaks in the power spectrum with a quadratic subtraction of the mean noise level. 
The analysis of the 184 \textsc{Elodie} spectra in addition to the \textsc{Coralie} spectra allows us to diminish the
first and second daily aliases (11.57 and 23.14\,$\mu$Hz) by  only 9\,\%. Note that a full coverage of \textsc{Elodie} data would
have allowed us to diminish daily aliases by 33\,\%.
\subsection{Search for a comb-like pattern}
\label{clp}
\begin{figure}
\resizebox{\hsize}{!}{\includegraphics{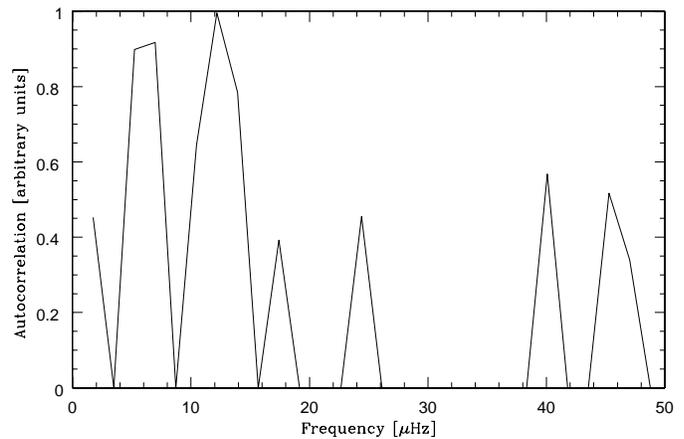}}
\caption{Autocorrelation of the power spectrum undersampled with a resolution of
1.75\,$\mu$Hz and a threshold of 0.2\,m$^2$\,s$^{-1}$. The large splitting is estimated
to about 40\,$\mu$Hz.}
\label{fig:ac}
\end{figure}
In solar-like stars, p-mode oscillations of low-degree are expected to produce a
characteristic comb-like structure in the power spectrum with mode
frequencies
$\nu_{n,l}$ reasonably well approximated by the asymptotic
relation (Tassoul \cite{tassoul80}):
\begin{eqnarray}
\label{eq1}
\nu_{n,l} & \approx &
\Delta\nu(n+\frac{l}{2}+\epsilon)-l(l+1) D_{0}\;.
\end{eqnarray}
Here, $D_0$, which is equal to $\frac{1}{6} \delta\nu_{02}$ if the 
asymptotic relation holds exactly, and $\epsilon$ are sensitive to the sound speed near the core and to
the surface layers respectively. 
The quantum numbers $n$ and $l$ correspond to the radial
order and the angular degree of the modes, and $\Delta\nu$ and
$\delta\nu_{02}$
to the large and small separations.
To search for periodicity in the power spectrum, an autocorrelation
is calculated and presented in Fig.~\ref{fig:ac}. As the large spacing is not
strictly constant, we use an undersampled power spectrum with a resolution of 1.75\,$\mu$Hz.
We are thus less sensitive to small variations of the large spacing versus the frequency.
Each peak of the autocorrelation corresponds to a structure present in the power spectrum.
The two strong peaks at low frequency close to 11.5 and 23\,$\mu$Hz
correspond to the daily aliases. The most probable large spacing
is situated near 40\,$\mu$Hz for \object{$\eta$ Boo}.
The result of the autocorrelation confirms the large spacing of 40.4\,$\mu$Hz deduced
from Kjeldsen et al. (\cite{kjel2}).

\subsection{Mode identification}

\begin{table}
\caption[]{Identification of extracted frequencies. Some frequencies can be either $\ell=1$ modes or due to noise.}
\begin{center}
\begin{tabular}{rcc}
\hline
\hline
\multicolumn{1}{c}{Frequency} & Mode ID  & S/N \\
\multicolumn{1}{c}{$[\mu$Hz$]$} &           &     \\
\hline
     512.2  &  $\ell = 1$   & 3.2 \\
$544.6 - 11.6 = 533.0$  &  $\ell = 0$   & 3.0 \\
     550.3  &  $\ell = 1$   & 3.0 \\
     589.9  &  $\ell = 1$   & 4.6 \\
$622.2 - 11.6 = 610.6$  &  $\ell = 0$   & 3.6 \\
$614.1 + 11.6 = 625.7$  &  $\ell = 1$   & 3.3 \\
$653.8 + 11.6 = 665.4$  &  $\ell = 1$   & 3.3 \\
     669.9  &  $\ell = 1$   & 3.9 \\
     691.3  &  $\ell = 0$   & 4.4 \\
     724.5  &  $\ell = 2$   & 5.2 \\
     728.3  &  noise   & 3.4 \\
     729.5  &  $\ell = 0$   & 4.6 \\
     748.5  &  $\ell = 1$   & 6.5 \\
$777.2 - 11.6 = 765.6$  &  $\ell = 2$   & 5.0 \\
$781.0 - 11.6 = 769.4$  &  $\ell = 0$   & 4.1 \\
$775.8 + 11.6 = 787.4$  &  $\ell = 1$   & 3.3 \\
     805.1  &  $\ell = 2$   & 3.0 \\
     809.2  &  $\ell = 0$   & 3.7 \\
$834.5 + 11.6 = 846.1$  &  $\ell = 2$   & 4.1 \\
     888.7  &  $\ell = 2$   & 4.0 \\
     891.6  &  $\ell = 0$   & 3.2 \\
     947.6  &  $\ell = 1$   & 3.1 \\
$960.3 + 11.6 = 971.9$  &  $\ell = 0$   & 3.7 \\
\hline
\end{tabular}
\end{center}
\label{tab:identif}
\end{table}
\begin{figure}[thb]
\resizebox{\hsize}{!}{\includegraphics{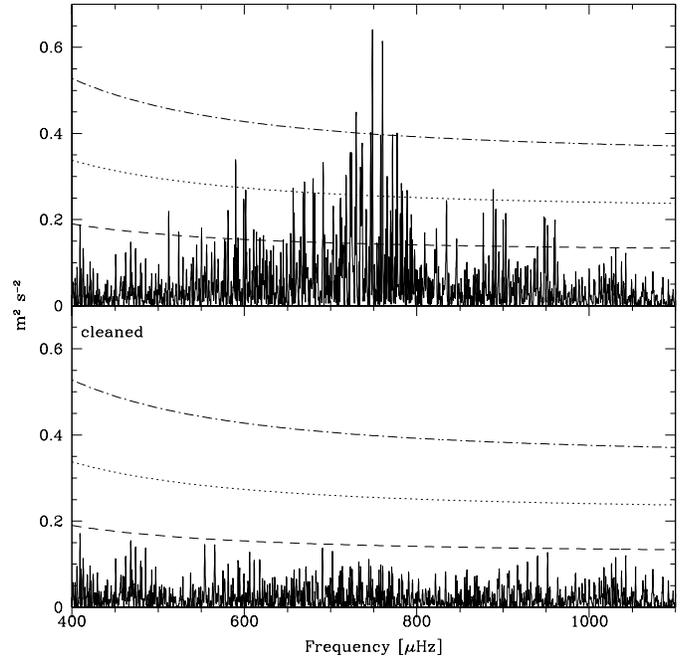}}
\caption[]{{\bf Top:} Original power spectrum of \object{$\eta$ Boo}. {\bf Bottom:} Cleaned power spectrum: all peaks listed
in Table~\ref{tab:identif} have been removed. The dot-dashed, dotted and dashed lines indicate an amplitude of 5\,$\sigma$, 
4\,$\sigma$ and 3\,$\sigma$, respectively. Numerous peaks are still present below 3\,$\sigma$, 
since no peaks have been cleaned below this threshold. These peaks can be due to p--mode oscillations and noise or have 
artificially been
added by the extraction algorithm due to the finite lifetimes of the modes} 
\label{clean}
\end{figure}

\begin{figure*}[thb]
\begin{center}
\resizebox{\hsize}{!}{\includegraphics{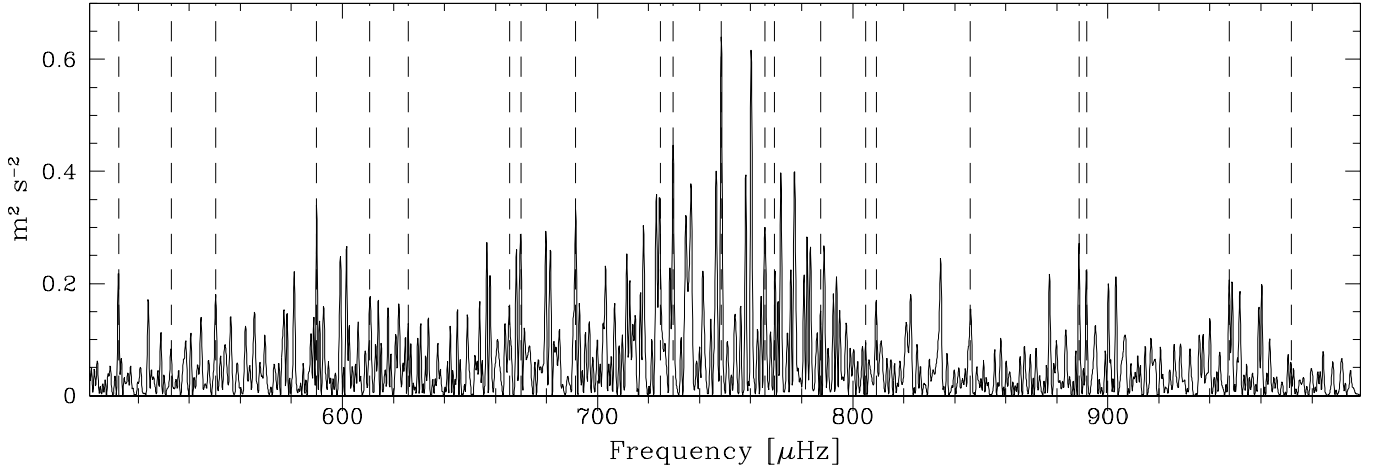}}
\caption[]{Power spectrum of \object{$\eta$ Boo} with the twenty-two extracted frequencies indicated by dashed lines.
The identification of each extracted frequency is given in Table~\ref{tab:freq}.}
\label{tfiden}
\end{center}
\end{figure*}

The frequencies were extracted using an iterative algorithm, which identifies the highest peak between
400 and 1100\,$\mu$Hz and subtracts
it from the time series. 
First, we iterated this process until all peaks with an amplitude higher than $3\sigma$ were removed 
(see Fig. \ref{clean}). $\sigma$ represents the noise in the amplitude spectrum decreasing from 14.5 to 12.2\,cm\,s$^{-1}$
in the above mentioned interval (see Sect.~\ref{asa}). Peaks with amplitudes below the 3$\sigma$ threshold were 
not considered since they
were too strongly influenced by noise and by interactions between noise and daily aliases.
This threshold, which ensures that the selected peaks have
only a small
chance to be due to noise, gave a total of twenty-three frequencies (see Table~\ref{tab:identif}).
Because of the daily alias of 11.57 $\mu$Hz, 
we cannot know {\it a priori} whether the frequency
selected by the algorithm is the right one or an alias. We thus considered
that the frequencies could be shifted by $\pm 11.57$ $\mu$Hz, and made echelle diagrams
for different large spacings near 40\,$\mu$Hz until every frequency could be identified as
an $\ell=0$, $\ell=1$ or $\ell=2$ mode. 
In this way, we found an averaged large spacing of 39.9 $\mu$Hz.   
It is difficult to identify $\ell=1$ modes as they appear to be mixed modes. Some identified $\ell=1$ modes
could thus be rather due to noise, e.g. peaks at 625.7 and 665.4\,$\mu$Hz.
The peak at 728.3\,$\mu$Hz was attributed to the noise, however it could be related to the $\ell=0$ mode 729.5\,$\mu$Hz
"split" owing to its lifetime.

\begin{figure}
\resizebox{\hsize}{!}{\includegraphics{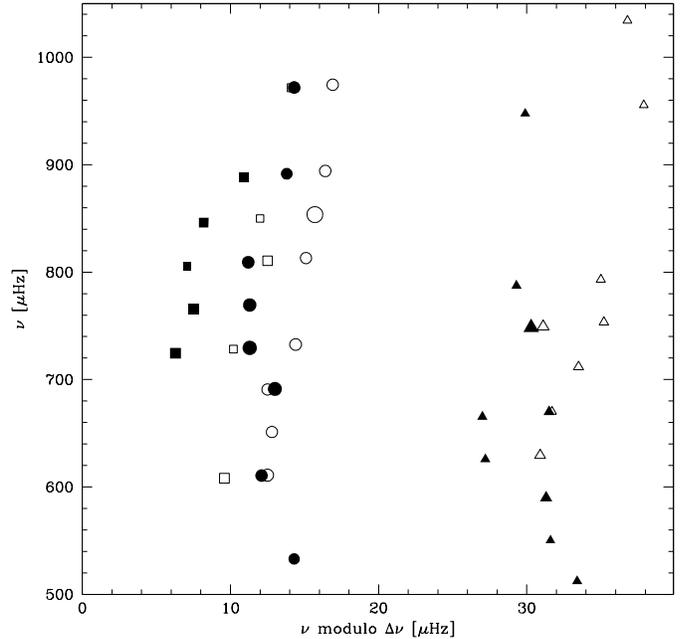}}
\caption{Echelle diagram of identified modes (in black) with a
large separation of 39.9\,$\mu$Hz. The
modes $\ell$=2 ($\blacksquare$), $\ell$=0 ({\Large $\bullet$}),
and $\ell$=1 ($\blacktriangle$) 
are represented with a size proportional to
their amplitude. Open symbols correspond to modes determined
by Kjeldsen et al. (\cite{kjel2}). Both identifications are
in good agreement at low frequency. However, at high frequency
their $\ell=2$ mode frequencies($\square$) seem to be identified in this paper
as $\ell=0$ modes 
({\Large $\bullet$}), and their
$\ell=0$ modes ({\tiny $\bigcirc$}) are not present in our data.}
\label{fig:ed}
\end{figure}

\begin{table}
\caption{Oscillation frequencies (in $\mu$Hz). The frequency resolution
of the time series is 0.87\,$\mu$Hz.}
\begin{center}
\begin{tabular}{lccc}
\hline
\hline
 & $\ell$ = 0 & $\ell$ = 1 & $\ell$ = 2 \\
\hline
n = 11 &  	& 512.2  &			\\
n = 12 & 533.0 	& 550.3	 &			\\
n = 13 &        & 589.9	 &			\\
n = 14 & 610.6  & 625.7  &    \\
n = 15 &        & 665.4 / 669.9	 &	  \\
n = 16 & 691.3  &        &724.5  \\
n = 17 & 729.5  &748.5  & 765.6	  \\
n = 18 & 769.4  & 787.4  &805.1  \\
n = 19 & 809.2  &		       &	846.1	      \\
n = 20 &  &		       &	888.7	      \\
n = 21 &891.6  &		       &		      \\
n = 22 &  &	947.6	       &		      \\
n = 23 &971.9 &		       &		      \\
\hline
$\Delta\nu_{\ell}$ & 39.9 (0.1) & 39.7 (0.2)& 40.9 (0.2)\\
\hline
\end{tabular}\\
\end{center}
\label{tab:freq}
\end{table}
The echelle diagram showing the twenty-two identified modes is shown in Fig.~\ref{fig:ed}. 
The frequencies of the modes are shown in Fig.~\ref{tfiden} and are given in Table~\ref{tab:freq}, 
with radial order of each oscillation mode deduced from the asymptotic relation 
(see Eq.~1) assuming that the parameter $\epsilon$ is near the solar value 
($\epsilon_{\odot} \sim 1.5$).
We can see that the oscillation modes do not strictly follow the asymptotic relation due to mixed $\ell=1$ modes
and a large curvature of others modes in the echelle diagram.
The average small spacing has a value of $\delta\nu_{02} = 3.95 \pm 0.9$\,$\mu$Hz.
The large spacing is separately determined for each value of $\ell$ and is
given in the last line of Table~\ref{tab:freq}. The weighted average of these three 
$\Delta\nu_{\ell}$ yields the value of $\Delta\nu = 39.9 \pm 0.1$\,$\mu$Hz.

The twenty-two identified modes are compared to previously identified ones by Kjeldsen et al. (\cite{kjel2}) in Fig.~\ref{fig:ed}.
Both identifications are rather in good agreement at low frequency but present major discrepancies at high frequency.
Although the present data have a higher $S/N$, additional measurements are needed to resolve
these ambiguities.

\subsection{Oscillation amplitudes}
\label{sec amp}
Concerning the amplitudes of the modes, theoretical computations predict oscillation 
amplitudes between 1 and 1.5\,m\,s$^{-1}$ for a 1.6 $M_{\odot}$ star 
like \object{$\eta$ Boo}, with mode lifetimes of the order of a few days
(Houdek et al. \cite{ho99}). The amplitudes of the highest modes, 
in the range 55--80\,cm\,s$^{-1}$, are then lower than expected. The observations indicate that oscillation amplitudes
are typically 2.5--3.5 times solar. 
This disagreement can be partly explained by the lifetimes of the modes. 
Indeed, the oscillation modes have finite lifetimes,
because they are continuously damped. Thus, if the star is observed during a time longer than 
the lifetimes of the modes, the signal is weakened due to
the damping of the modes and to their re--excitation with a random phase.

\section{Comparison with models}
\label{mod}
In order to compare the asteroseismic observations with theoretical predictions, 
stellar models were computed using the Geneva evolution code including shellular rotation
and atomic diffusion (see Eggenberger et al. \cite{eg04c} for more details). 
We used the OPAL opacities, the NACRE nuclear reaction rates 
(Angulo et al. \cite{an99}) and the standard mixing--length formalism
for convection.

\subsection{Non--asteroseismic observational constraints}
\label{obs}

Following Di Mauro et al. (\cite{di03}) (hereafter DM03), 
the luminosity of \object{$\eta$ Boo} was deduced from the Hipparcos parallax:
$L/L_{\odot}=9.02 \pm 0.22$.
Concerning the effective temperature, 
we added recent results to the references used by DM03 to determine
a new average value (see Table~\ref{tab:temp}). As a result, the effective temperature
$T_{\mathrm{eff}}=6030 \pm 90$\,K was adopted. This value is in perfect agreement with the
effective temperature of $6028 \pm 45$\,K used by DM03, with a larger error which seems
to us more realistic in view of the different values found in the literature. 
The box in the HR diagram for \object{$\eta$ Boo} 
delimited by the observed effective temperature and luminosity is shown in Fig.~\ref{dhr}.

The metallicity of \object{$\eta$ Boo} adopted by DM03 is [Fe/H$]=0.305 \pm 0.051$.
Compared to different observed metallicities, this value seems to be quite large. 
We thus decided to adopt
a lower average value of [Fe/H$]=0.23 \pm 0.07$, which is determined from the recent measurements
listed in Table~\ref{tab:temp}.    

Finally, we used the observed surface velocity of \object{$\eta$ Boo} to constrain the rotational velocity of our models.
From Coralie spectra, we determined a rotational velocity $v \sin i=12.8$\,km\,s$^{-1}$. Since the value of the 
angle $i$ is unknown, we assumed that it is close to 90\,$\degr$. Thus our models of \object{$\eta$ Boo} have to reproduce a
surface velocity of about 13\,km\,s$^{-1}$. 

\begin{table}
\caption{Metallicity and temperature determination for \object{$\eta$ Boo} (since 1990). 
The errors on the selected parameters are
chosen to encompass all acceptable values (last line).}
\begin{center}
\begin{tabular}{ll|l}
\hline
\hline
 $T_{\rm eff}$ [$\degr$ K] & [$Fe/H$] & References \\
\hline
6000 & 0.16 & McWilliam \cite{mcw}\\
6219 & 0.30/0.37& Balachandran \cite{bal}\\
6068 & 0.19 & Edvardsson et al. \cite{edv}\\
5943 & 0.20 & Gratton et al. \cite{gra}\\
--   & 0.23 & Mishenina \cite{mis}\\
--   & 0.28 & Fuhrma (Cayrel de Strobel \cite{cay})\\
6003 & 0.25 & Cenarro et al. \cite{cen}\\
--   & 0.16 & Buzzoni et al. \cite{buz}\\
6000 & 0.25 & Feltzing \& Gonzalez \cite{fel}\\
6120 & 0.16& Gray \cite{gray}\\
5964 & 0.19& Mallik et al. \cite{mal}\\
5957 & 0.15 & Nordstrom et al. \cite{nord} \\
5942 & 0.18 & Allende Prieto et al. \cite{all}\\
\hline
6030\,$\pm$\,90 & 0.23\,$\pm$\,0.07 & \\
\hline
\end{tabular}\\
\end{center}
\label{tab:temp}
\end{table}

\subsection{Computational method}

Basically, the computation of a stellar model for a given star consists in finding the set of stellar modeling parameters which best reproduces 
all observational 
data available for this star.
The characteristics of a stellar model including the effects of rotation 
depend on six modeling parameters: the mass $M$ of the star, its age ($t$ hereafter), 
the mixing--length parameter $\alpha \equiv l/H_{\mathrm{p}}$ for convection, the initial surface velocity $V_{\mathrm{i}}$
and two parameters describing the initial chemical composition of the star. For these two parameters,
we chose the initial hydrogen abundance $X_{\mathrm{i}}$ and the initial ratio between the mass fraction of heavy elements and hydrogen 
$(Z/X)_{\mathrm{i}}$. 
Assuming that this ratio is proportional to the abundance ratio [Fe/H], we can directly relate $(Z/X)$ to [Fe/H]
by using the solar value $(Z/X)_{\odot}=0.0230$ given by Grevesse \& Sauval (\cite{gr98}).
Moreover, we fixed the mixing--length parameter to its solar calibrated value ($\alpha_{\odot}=1.75$) and we assumed the  
initial hydrogen abundance $X_{\mathrm{i}}$ to be $X_{\mathrm{i}}=0.7$.
As a result, any characteristic $A$ of a given stellar model has the following formal dependences with respect to modeling parameters:
$A=A(M,t,V_{\mathrm{i}},(Z/X)_{\mathrm{i}})$.\\

\begin{figure}[htb!]
 \resizebox{\hsize}{!}{\includegraphics{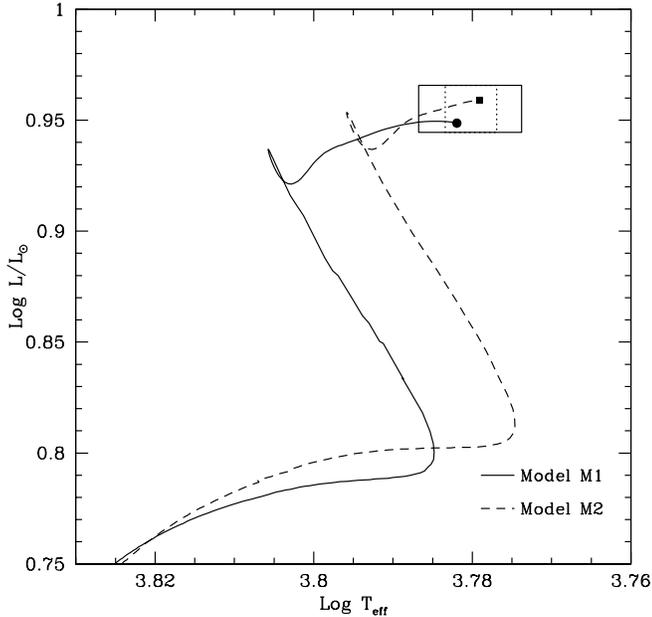}}
  \caption{Evolutionary tracks in the HR diagram for two models of \object{$\eta$ Boo}.
The dot and the square indicate the location of the M1 and M2 model respectively.
The error box in continuous line indicates the observational constraints of $L$ and $T_{\mathrm{eff}}$ used in our
analysis, while the dotted box corresponds to the constraints used by Di Mauro et al. (\cite{di03}).}
  \label{dhr}
\end{figure}

The determination of the set of modeling parameters $(M,t,V_{\mathrm{i}},(Z/X)_{\mathrm{i}})$ leading to the best agreement
with the observational constraints is made in two steps. First, we constructed a grid of models with position in the HR diagram 
in agreement with the observational values of the luminosity and effective temperature (see Fig.~\ref{dhr}).
Note that the initial ratio between the mass fraction of heavy elements and hydrogen 
$(Z/X)_{\mathrm{i}}$ is directly constrained by the observed surface metallicity [Fe/H], while the initial velocity
$V_{\mathrm{i}}$ is directly constrained by the observed rotational velocity.

For each stellar model of this grid, low-$\ell$ p--mode frequencies were then calculated using the Aarhus adiabatic 
pulsations package written by J. Christensen-Dalsgaard (\cite{cd97}). Following our observations, modes $\ell \leq 2$
with frequencies between 0.4 and 1.1\,mHz were computed and the mean large ($\Delta \nu$) and small spacings ($\delta \nu_{02}$)
were determined. The mean large spacing was computed by considering only the radial modes. 
Once the asteroseismic characteristics of all models of the grid were determined, we performed a $\chi^2$ minimization
as in Eggenberger et al. (\cite{eg04b}). Thus, two functionals are defined: $\chi^2_{\mathrm{tot}}$ and $\chi^2_{\mathrm{astero}}$.
The $\chi^2_{\mathrm{tot}}$ functional is defined as follows:
\begin{eqnarray}
\label{eq11}
\chi^2_{\mathrm{tot}} \equiv \sum_{i=1}^{5} \left( \frac{C_i^{\mathrm{theo}}-C_i^{\mathrm{obs}}}{\sigma C_i^{\mathrm{obs}}} \right)^2  \; ,
\end{eqnarray}
where the vectors $\mathbf{C}$ contains the following observables for one star:  
\begin{eqnarray}
\nonumber
\mathbf{C} \equiv (L/L_{\odot},T_{\mathrm{eff}},
[\mathrm{Fe/H}],\Delta \nu,\delta \nu_{02}) \; .   
\end{eqnarray} 
The vector $\mathbf{C}^{\mathrm{theo}}$ contains the theoretical values of these observables for the model to be tested, while 
the values of $\mathbf{C}^{\mathrm{obs}}$ are those
listed above. The vector $\mathbf{\sigma C}$ contains the errors on these observations.
Note that the observed rotational velocity is not included in this minimization, because of its large uncertainty resulting
from the unknown inclination angle $i$.
The $\chi^2_{\mathrm{astero}}$ functional is defined as follows:
\begin{eqnarray}
\nonumber
\chi^2_{\mathrm{astero}} & \equiv & \frac{1}{N} \sum_{i=1}^{N} \left(
\frac{\nu_i^{\mathrm{theo}}-\nu_i^{\mathrm{obs}} - \langle D_{\nu}\rangle}{\sigma} \right)^2\,, \\
\label{eq2}
\end{eqnarray} 
where $\sigma=0.87$\,$\mu$Hz is the error on the observed frequencies estimated as the frequency resolution, $N=22$
is the number of observed frequencies, and $\langle D_{\nu}\rangle$ is the mean value of the differences between
the theoretical and observed frequencies :
\begin{eqnarray}
\nonumber
\langle D_{\nu}\rangle \equiv \frac{1}{N} \sum_{i=1}^N (\nu_i^{\mathrm{theo}}-\nu_i^{\mathrm{obs}}) \; .
\end{eqnarray}
The determination of the best set of parameters
was based on the minimization of the functional defined in equation~(\ref{eq11}) which includes three non--asteroseismic
and two asteroseismic observational constraints. 
Once the model with the smallest $\chi^2_{\mathrm{tot}}$ was determined, we refined the grid in the vicinity of this
preliminary solution in order to find the best solution which minimizes at the same time $\chi^2_{\mathrm{tot}}$ and $\chi^2_{\mathrm{astero}}$.

\subsection{Results}

Using the observational constraints listed in Sect.~\ref{obs} with the observed frequencies listed in Table~\ref{tab:freq},
we performed the $\chi^2$ minimization described above. We found the solution $M=1.57 \pm 0.07$\,$M_{\odot}$,
$t=2.67 \pm 0.10$\,Gyr, $V_{\mathrm{i}}\cong 90$\,km\,s$^{-1}$ and $(Z/X)_{\mathrm{i}}=0.0391 \pm 0.0070$. 
The position of this model in the HR
diagram (denoted model M1 in the following) is indicated by a dot in Fig.~\ref{dhr}. 
The characteristics of this model are given in Table~\ref{tab:res}.
The confidence limits of each modeling parameter given in Table~\ref{tab:res} are estimated as the maximum/minimum values which 
fit the observational constraints when the other calibration parameters are fixed to their medium value.
Note that the radius deduced for \object{$\eta$ Boo} $R\,=\,2.72\,R_{\odot}$ is in good agreement with the
interferometric radius of 2.766$\pm$0.039\,$R_{\odot}$ and 2.682$\pm$0.043\,$R_{\odot}$ determined respectively 
with the Mark III optical interferometer  (Mozurkewich et al. \cite{mozur}) and 
with the VLTI (Kervella et al. \cite{kerv}).

\begin{figure}[htb!]
 \resizebox{\hsize}{!}{\includegraphics{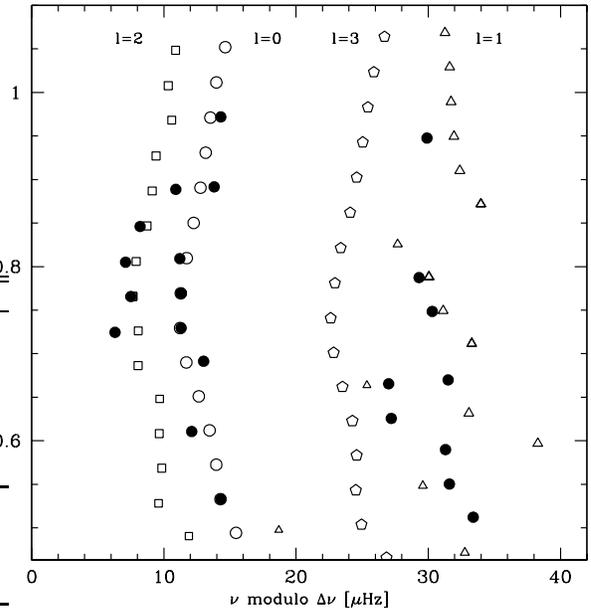}}
  \caption{Echelle diagram for the M1 model with a large spacing $\Delta \nu=39.9$ $\mu$Hz. Open symbols refer to theoretical
  frequencies, while the filled circles correspond to the observed frequencies
  listed in Table~\ref{tab:freq}. Open circles are used for modes with $\ell=0$, triangles for $\ell=1$, squares for $\ell=2$ and pentagons for $\ell=3$. The
  size of the open symbols is proportional to the relative surface amplitude of the mode (see text). Modes with too small
  surface amplitudes (e.g. g modes) are not shown on this diagram.}
  \label{figdiagech}
\end{figure}

Theoretical p--mode frequencies of the M1 model are compared to the observed frequencies by plotting the echelle diagram (see Fig.~\ref{figdiagech}).
Note that in this figure the systematic difference $\langle D_{\nu}\rangle$ between theoretical and observed frequencies
has been taken into account. Following Christensen-Dalsgaard et al. (\cite{cd95}), the theoretical oscillation amplitudes are estimated by 
\begin{equation}  
\frac{A_{nl}}{A_0(\nu_{nl})} \cong \sqrt{\frac{\epsilon_{0}(\nu_{nl})}{\epsilon_{nl}}} \quad ,
\end{equation}
where $A$ is the surface amplitude and $\epsilon$ the normalized energy of the mode. $A_0(\nu)$ and $\epsilon_0(\nu)$ are
obtained by interpolating to frequency $\nu$ in the results of radial modes.

The model shows that $\ell=1$ modes deviate from the asymptotic relation for p--modes. 
This is a consequence of the avoided crossings (Christensen--Dalsgaard et al. \cite{cd95}; Guenther \& Demarque \cite{gu96}; 
DM03; Guenther \cite{gu04}, hereafter GU04).
This results in frequencies which are shifted relative to the frequencies expected for pure p--modes. 
This is particularly true for the $\ell=1$ modes at low frequency which strongly deviate from the asymptotic relation.
The observed frequencies for $\ell=1$ modes seem to deviate from the asymptotic relation, which is in accordance with
theoretical predictions. However, Fig. \ref{figdiagech} shows that the model results are not able to precisely reproduce
the individual frequencies of these modes at low frequency. 

Figure~\ref{figdiagech} also shows that
the agreement between observations and theoretical predictions for modes with $\ell=0$ and $\ell=2$
is good, except for the two $\ell=2$ modes with
the smallest and the largest frequency ($\nu=724.5$ and $888.7$\,$\mu$Hz).
Note that the $\ell=2$ modes are also influenced by the avoided crossings.
However, the effects of coupling
become much weaker for these modes than for modes with $\ell=1$, since p--modes with $\ell=2$ penetrate less deep in the stellar interior.

\begin{table}
\caption[]{Models for \object{$\eta$ Boo}. The M1 and M2 models include rotation and atomic diffusion,
while the M3 model is a standard model computed with an overshooting parameter $\alpha_{\mathrm{ov}}=0.2$.
The upper part of the table gives the observational constraints used for the
calibration. The middle part of the table presents the modeling parameters with their confidence limits, while the bottom
part presents the global parameters of the star.}
\begin{center}
\label{tab:res}
\begin{tabular}{c|ccc}
\hline
\hline
 & Model M1 & Model M2 & Model M3\\ \hline
$L/L_{\odot}$ & $9.02 \pm 0.22$ & $9.02 \pm 0.22$ & $9.02 \pm 0.22$ \\
$T_{\mathrm{eff}}$ [K]& $6030 \pm 90$ & $6028 \pm 45$ & $6030 \pm 90$ \\
$(Z/X)_{\mathrm{s}}$ & $0.039 \pm 0.007$ & $0.057 \pm 0.007$ & $0.039 \pm 0.007$ \\
$V$ [km\,s$^{-1}$] & $\sim 12.8$ & $\sim 12.8$ & -\\
$\Delta \nu$ [$\mu$Hz] & $39.9 \pm 0.1$  & $39.9 \pm 0.1$ & $39.9 \pm 0.1$ \\
$\delta \nu_{02}$ [$\mu$Hz] & $3.95 \pm 0.90 $ & $3.95 \pm 0.90$ & $3.95 \pm 0.90 $\\ 
\hline
$M$ $[M_{\odot}]$ &  $1.57 \pm 0.07$ & $1.69 \pm 0.05$ & $1.70 \pm 0.05$\\
$t$ [Gyr] &  $2.67 \pm 0.10$ & $2.65 \pm 0.10$ & $2.14 \pm 0.10$ \\
$V_{\mathrm{i}}$ [km\,s$^{-1}$] & $\sim 90 $& $\sim 80$ & -\\
$(Z/X)_{\mathrm{i}}$ & $0.0391 \pm 0.0070$ & $0.0580 \pm 0.0070$ & $0.0390 \pm 0.0070$   \\
\hline
$L/L_{\odot}$ & $8.89$ & $9.10$ & $9.11$\\
$T_{\mathrm{eff}}$ [K]& $6053$ & $6013$ & $6005$\\
$R/R_{\odot}$ & $2.72$ & $2.78$ & $2.79$\\
$V$ [km\,s$^{-1}$] & $12$ & $15$ & -\\
$(Z/X)_{\mathrm{s}}$ & $0.0388$ & $0.0570$ & $0.0390$\\
$\Delta \nu$ [$\mu$Hz] & $39.9$  & $39.9$ & $39.9$\\
$\delta \nu_{02}$ [$\mu$Hz] & $3.79$ & $3.67$ & $3.45$\\
\hline
\end{tabular}
\end{center}
\end{table}
  
\begin{figure}[htb!]
 \resizebox{\hsize}{!}{\includegraphics{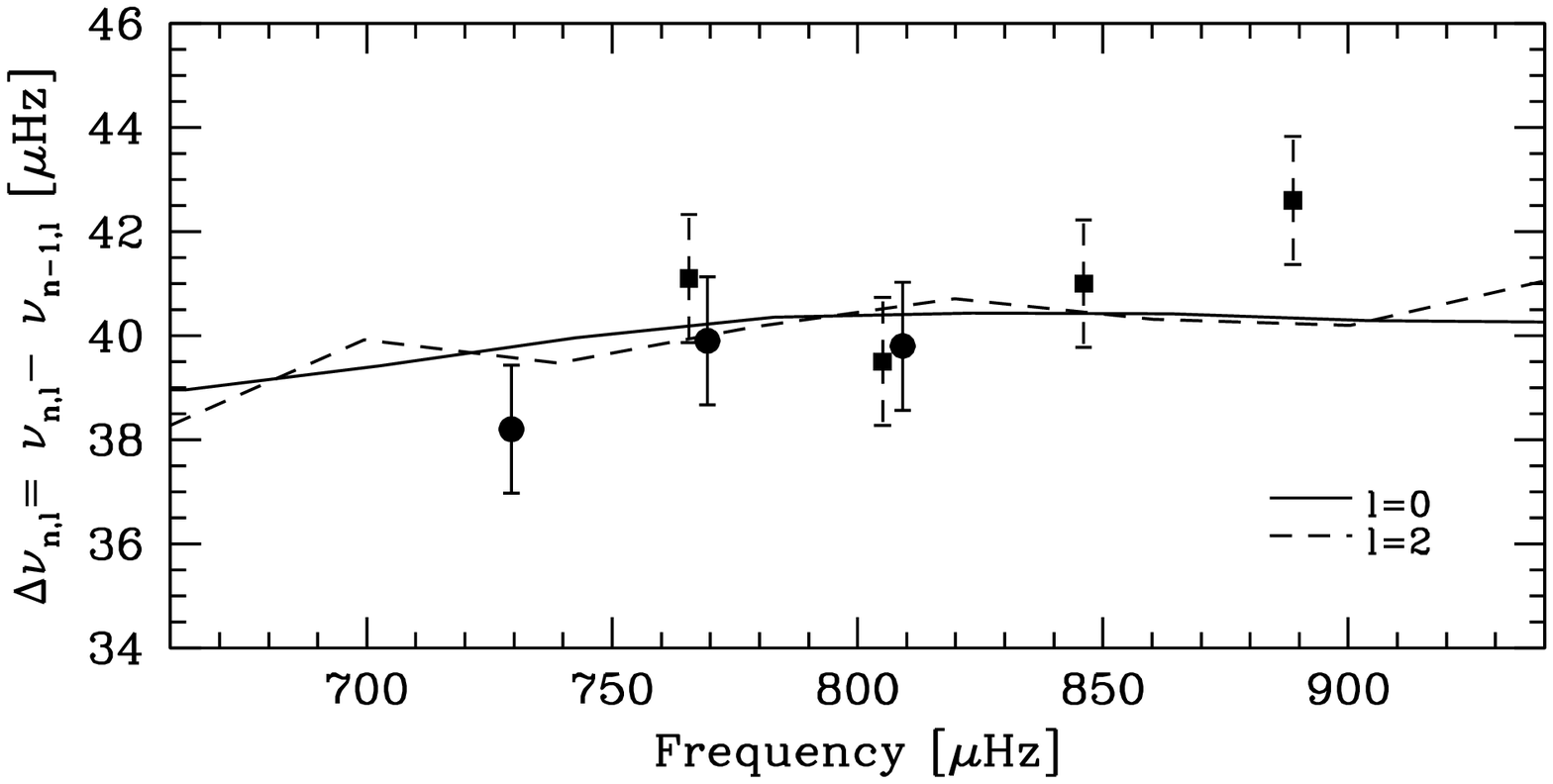}}
 \resizebox{\hsize}{!}{\includegraphics{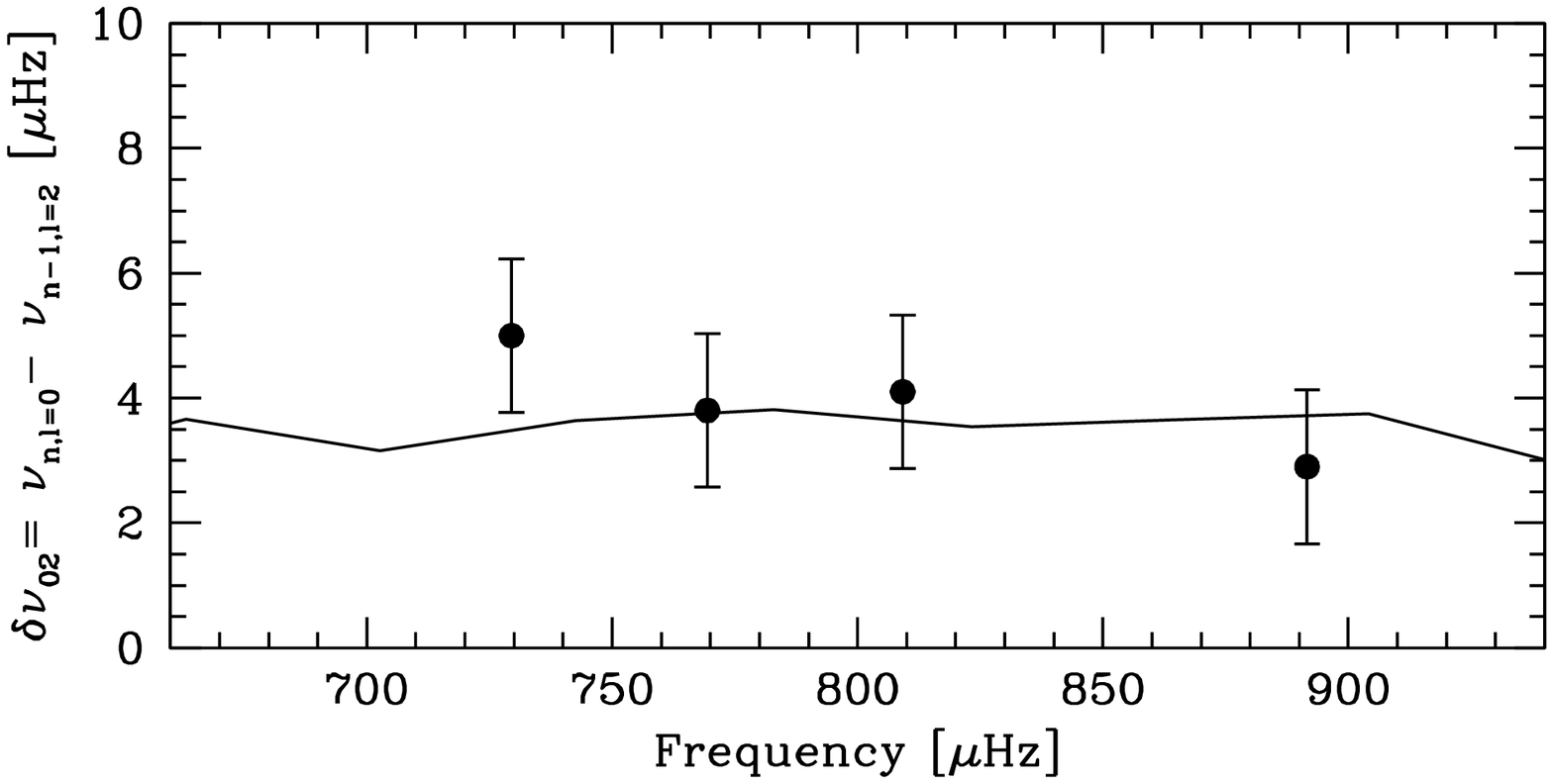}}
   \caption{Large and small spacings versus frequency for the M1 model.
  Dots indicate the observed values of the large ($\ell=0$) and small spacings, while squares correspond
  to the large spacing determined with $\ell=2$ modes.}
  \label{gdpt}
\end{figure}

The variations of the large and small spacing with the frequency are given in Fig.~\ref{gdpt}.
Large spacings for $\ell=1$ modes are not plotted in Fig.~\ref{gdpt}, since these modes deviate
too strongly from the asymptotic behaviour of pure p--modes.
Table~\ref{tab:res} and Fig.~\ref{gdpt} show that the mean large spacing of the M1 model is in perfect
agreement with the observed value. The observed variation of the large spacing with the frequency is also correctly reproduced
by the model, except for the large value of the $\ell=2$ point close to 900\,$\mu$Hz. 
Table~\ref{tab:res} and Fig.~\ref{gdpt} also show that the observed small spacings are compatible with theoretical predictions.
The observed mean small spacing is however slightly larger than the theoretical one; this is mainly due to the large value 
of the observed small spacing at $724.5$\,$\mu$Hz.  
    
We conclude that the observed frequencies listed in Table~\ref{tab:freq} are compatible with theoretical predictions. 
Although the general agreement is satisfactory,
we also note some discrepancies between observed and predicted frequencies, especially for the $\ell=1$ modes 
at low frequency. 

\subsection{Discussion of the results and comparison with previous studies}

Detailed studies of \object{$\eta$ Boo} based on the asteroseismic observations of
Kjeldsen et al. (\cite{kjel2}) have already been
performed by DM03 and GU04. 
Compared to these studies, we notice that our M1 model has a smaller mass. Indeed, DM03 found that the mass of \object{$\eta$ Boo}
is limited to the range $M=(1.64-1.75)$\,$M_{\odot}$, while GU04 proposed two different solutions: a model with a mass
of 1.706\,$M_{\odot}$ which has exhausted its hydrogen core,
and another model with a mass of 1.884\,$M_{\odot}$ which is still on the main--sequence, but is approaching hydrogen exhaustion.
These two authors used the same non--asteroseismic
constraints (see Sect.~\ref{obs}). However, contrary to the analysis by DM03 and contrary to the present work, GU04 used
a calibration method (the QDG method) which is not limited to models with position in the HR diagram 
in agreement with the observational values of the luminosity and effective temperature. 
This explains why GU04 found another solution with a mass of 1.884\,$M_{\odot}$, while DM03 determined a mass
between 1.64 and 1.75\,$M_{\odot}$.
Recently, Di Mauro et al. (\cite{di04}) (hereafter DM04) showed that main--sequence models
provide a match to the observed location of \object{$\eta$ Boo} in the
HR diagram when overshooting from the convective core ($\alpha_{\mathrm{ov}} \geq 0.1$)
is included in the computation.

\begin{figure}[htb!]
 \resizebox{\hsize}{!}{\includegraphics{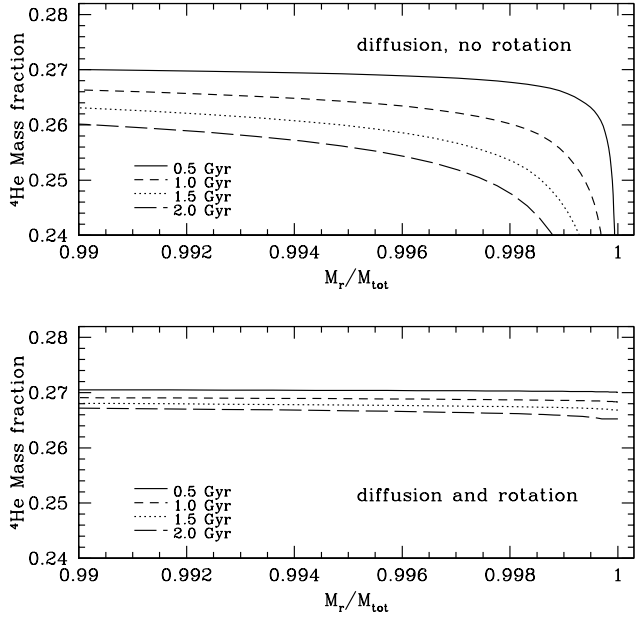}}
  \caption{Helium abundance profile in the external layers of the star 
  at different age during its evolution on the main--sequence. The model on top only includes
  atomic diffusion, while the model on bottom includes atomic diffusion and shellular rotation.
  Apart from the inclusion of rotation, the two models have been computed with the same initial
  parameters corresponding to the M1 model.  
  }
  \label{proy}
\end{figure} 
    
The fact that our M1 model is less massive than the solution of about 1.7\,$M_{\odot}$
found by DM03 and GU04 can either be due to the different observational constraints used or to the different
input physics of the evolution codes. 
Indeed, our models include shellular rotation and atomic diffusion contrary to the models calculated
by DM03 and GU04. For stars more massive than about 1.4\,$M_{\odot}$, 
it is necessary to introduce another transport mechanism,
like the rotationally induced mixing, in order to counteract the effect of atomic
diffusion in the external layers. When only atomic diffusion is included in
a star with a thin convective envelope, helium and heavy elements are drained out
of the envelope, resulting in too low surface abundances which are incompatible with 
observation. This is illustrated on Fig.~\ref{proy} which shows the helium profile in the external
layers at different age during the evolution on the main--sequence for a model
including only atomic diffusion and for the M1 model which includes shellular rotation and atomic
diffusion. Fig.~\ref{proy} shows that rotationally induced mixing prevents the helium from being 
drained out of the convective envelope. Indeed, the decrease of the surface helium abundance during the 
main--sequence evolution is found to be very small for models including rotation and atomic diffusion.

\subsubsection{Effect of the metallicity} 
 
To investigate the effects of non--asteroseismic observational constraints on
the solution, we decided to redo the whole calibration
using the non--asteroseismic constraints adopted by DM03 and GU04.
The metallicity was increased to $Z=0.04$
and a temperature of $T_{\mathrm{eff}}=6028 \pm 45$\,K was adopted. Note that we still used our asteroseismic constraints for this calibration.
In this way, we found the solution $M=1.69 \pm 0.05$\,$M_{\odot}$,
$t=2.65 \pm 0.10$\,Gyr, $V_{\mathrm{i}}\cong 80$\,km\,s$^{-1}$ and $(Z/X)_{\mathrm{i}}=0.0580 \pm 0.0070$.
The position of this model in the HR
diagram (noted model M2 in the following) is denoted by a square in Fig.~\ref{dhr}. 
The characteristics of the M2 model are given in Table~\ref{tab:res}.
We conclude that the difference in the adopted value for the metallicity explains the different mass determined. Indeed, Fig.~\ref{figdiagech}
shows that the fact that we used $T_{\mathrm{eff}}=6030 \pm 90$\,K instead of $T_{\mathrm{eff}}=6028 \pm 45$\,K
has no significant influence on the solution since the M1 model is also included in the smaller observational box determined by DM03.

The higher metallicity used by DM03 and GU04 results of course from the larger value of the observed [Fe/H]: they adopted 
[Fe/H$]=0.305 \pm 0.051$, while we fixed [Fe/H] to $0.23 \pm 0.07$ for the M1 calibration. However, we notice that it also results from the way 
one relates the observed [Fe/H] to the mass fractions $Z$ and $X$ used in the models. Indeed, we directly related $(Z/X)$ to [Fe/H]
by using the solar value $(Z/X)_{\odot}=0.0230$ given by Grevesse \& Sauval (\cite{gr98}), while DM03 related
$Z$, and not $(Z/X)$, to [Fe/H]. As a result, for a same value of [Fe/H$]=0.305$, we determined $Z=0.032$ while DM03 obtained
a higher value of $Z=0.040$ (for $X=0.7$).

We conclude that the derived mass of \object{$\eta$ Boo} is very sensitive to the choice of the observed metallicity. When
a metallicity of [Fe/H$]=0.23 \pm 0.07$ is adopted, a mass of $1.57 \pm 0.07$\,$M_{\odot}$ is found. When the higher metallicity determined by
DM03 is used, we obtain a mass of $1.69 \pm 0.05$\,$M_{\odot}$, in perfect agreement with the results of 
DM03 and GU04.

\begin{figure}[htb!]
 \resizebox{\hsize}{!}{\includegraphics{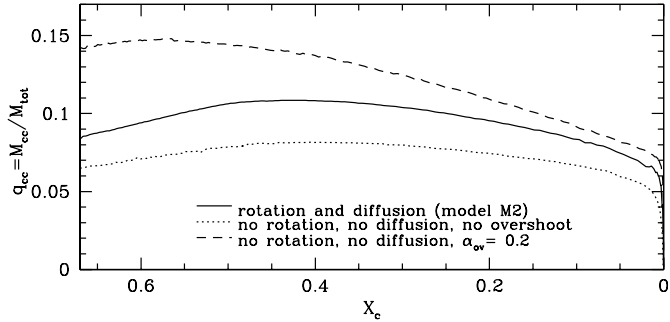}}
  \caption{Ratio of the mass of the convective core to the total mass of the star ($q_{\mathrm{cc}}$)
as a function of the central hydrogen abundance ($X_{\mathrm{c}}$), for a standard model without overshooting,
for a standard model with $\alpha_{\mathrm{ov}}=0.2$ and for a model with rotation and atomic diffusion.  
Apart from the inclusion of rotation and overshooting, the three models have been computed with the same initial
parameters corresponding to the M2 model.  
  }
  \label{ccore}
\end{figure} 
\subsubsection{Effect of the input physics}

Contrary to the masses, the ages of the M1 and M2 models are very similar (2.67 and 2.65\,Gyr respectively) and are therefore not very
sensitive to a change of metallicity. However, this age is larger than the age of $2.393$\,Gyr obtained by GU04 for its solution with a mass of 
$1.706$\,$M_{\odot}$. DM03 pointed out that the age of the models depends on the inclusion of overshooting: 
the age is about 2.3--2.4\,Gyr without overshooting, and between 2.4--2.7\,Gyr in presence of overshooting. The age of \object{$\eta$ Boo} 
seems therefore to be sensitive to the input physics used.
To investigate these effects on the solution, we decided to calculate models without rotation
and atomic diffusion using the same observational constraints as the M2 model. In this way, we find the solution
$M=1.70 \pm 0.05$\,$M_{\odot}$ and $t=2.39 \pm 0.10$\,Gyr, in perfect accordance with the results of GU04 and
DM03. Rotating models predict a larger age for \object{$\eta$ Boo} than non--rotating ones.
This illustrates the fact that, for small initial velocities, rotational effects are found to mimic the effects due
to an overshoot of the convective core into the radiative zone.  
Indeed, the lifetimes of rotating models are enhanced with respect to those of standard models, because
the mixing feeds the core with fresh hydrogen fuel. As a result, the exhaustion of hydrogen in the central
region is delayed and the time spent on the main-sequence increases. This can be seen on Fig.~\ref{ccore}
which shows the ratio of the mass of the convective core to the total mass of the star ($q_{\mathrm{cc}}$)
as a function of the central hydrogen abundance ($X_{\mathrm{c}}$).
We see that the rotating model exhibits a larger convective core for a given $q_{\mathrm{cc}}$, i.e. for
a given evolutionary stage on the main-sequence, than the standard model without overshooting. In the same
way, the non-rotating model with $\alpha_{\mathrm{ov}}=0.2$ also exhibits a larger convective core on the main-sequence
than standard models without overshooting. This explains why the inclusion of rotation or overshooting increases
the lifetimes of the model on the main-sequence and hence the deduced age for \object{$\eta$ Boo}.   
 
\begin{figure}[htb!]
 \resizebox{\hsize}{!}{\includegraphics{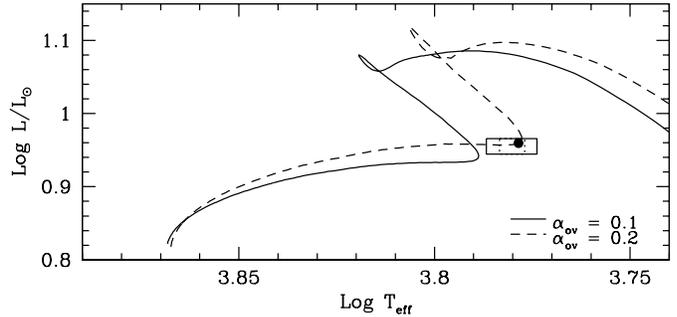}}
  \caption{Evolutionary tracks in the HR diagram for two models of 1.7\,$M_{\odot}$
  computed with a different amount of overshooting.
The dot indicates the location of the M3 model.
The error box in continuous line corresponds to the observational constraints of $L$ and $T_{\mathrm{eff}}$ used in our
analysis, while the dotted box corresponds to the constraints used by DM03.}
  \label{dhr_ov}
\end{figure} 

Finally, we investigated the solution of a model which is still on the main-sequence. As found by
DM04, these models do not provide a match to the observed $T_{\mathrm{eff}}$ and $L$ of \object{$\eta$ Boo}
unless overshooting is included. Thus, our analysis using the input physics described above leads
to only one solution which is in accordance with asteroseismic and non-asteroseismic
observables: the M1 model which is in the post-main-sequence phase of evolution.
Using the observational constraints listed in Sect.~\ref{obs}, we tried to determine a model of \object{$\eta$ Boo}
which is still on the main-sequence by computing non-rotating stellar models including overshooting.
In this way, we found that a model computed with an overshooting parameter $\alpha_{\mathrm{ov}}=0.2$ 
and a mass of 1.7\,$M_{\mathrm{\odot}}$ enables to match the location of \object{$\eta$ Boo} in the HR
diagram (see Fig.~\ref{dhr_ov}). As discussed above, the mass of this model (denoted model M3 in the following)
is lower than the mass of 1.884\,$M_{\odot}$ determined by GU04 and the
values of $M=(1.75-1.90)$\,$M_{\odot}$ found by DM04, because of the smaller metallicity used in our
analysis. The characteristics of the M3 model are given in Table~\ref{tab:res}.

\begin{figure}[htb!]
 \resizebox{\hsize}{!}{\includegraphics{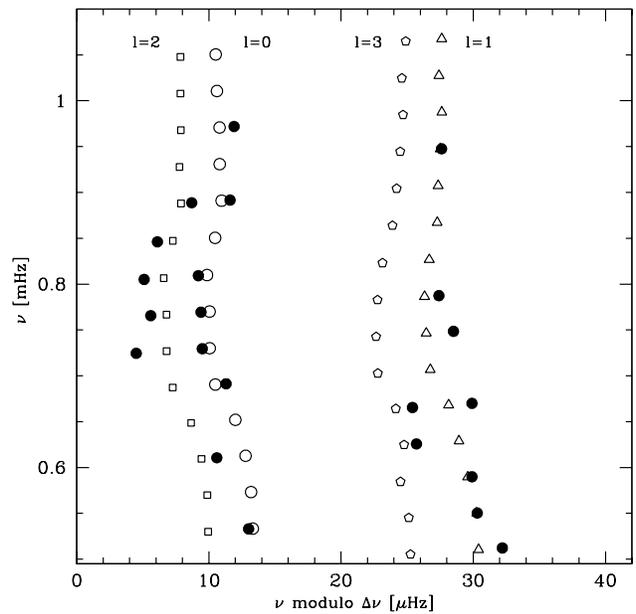}}
  \caption{Echelle diagram for the M3 model with a large spacing $\Delta \nu=39.9$ $\mu$Hz. Open symbols refer to theoretical
  frequencies, while the filled circles correspond to the observed frequencies
  listed in Table~\ref{tab:freq}. Open circles are used for modes with $\ell=0$, triangles for $\ell=1$, squares for $\ell=2$ and pentagons for $\ell=3$.}
  \label{diagech_M3}
\end{figure}

\begin{figure}[htb!]
 \resizebox{\hsize}{!}{\includegraphics{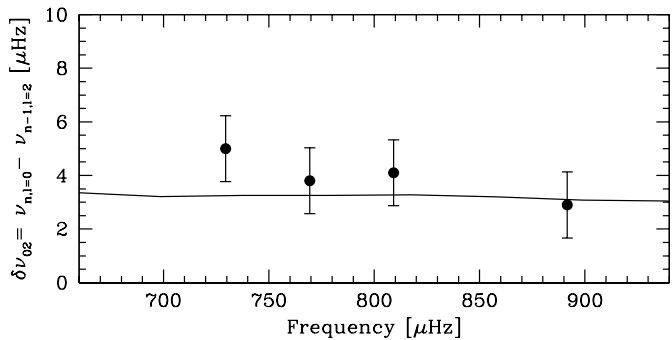}}
   \caption{Small separations versus frequency for the M3 model.
  Dots indicate the observed values of the small separations.}
  \label{pt_M3}
\end{figure} 
As already pointed out by GU04 and DM04, the fundamental seismic difference between
post-main-sequence (M1) and main-sequence (M3) models
concerns the avoided crossings.
Models in the post-main-sequence phase show $\ell=1$ modes that deviate
from the asymptotic relation, while models still on the main-sequence show 
no occurrence of avoided crossing. Indeed, for these models only modes with a radial order lower than
the observed ones are mixed (see Fig.~\ref{figdiagech} and \ref{diagech_M3}).               
Since observation show $\ell=1$ modes that deviate
from the asymptotic relation, we conclude that models in the
post-main-sequence phase of evolution are in better agreement with the asteroseismic measurements than the
main-sequence models. Moreover, the small separation of the M3 model is smaller than the one of the M1 model and is
therefore in slightly less good agreement with the observed frequencies (see Fig.~\ref{pt_M3}).
Note that DM04 also found that post-main-sequence models are characterized by larger small separations than
main-sequence models. Although the M1 model constitutes the solution 
which best reproduced all
observational constraints, the actual precision on the observed frequencies
does not enable us to definitively reject the solution on the main-sequence.

\section{Conclusions}
\label{conc}
Our observations of \object{$\eta$ Boo} yield a clear detection of p-mode oscillations.
Several identifiable modes appear in the power spectrum between 0.4 and 1.1\,$\mu$Hz
with average large and small spacings of 39.9 and 3.95\,$\mu$Hz respectively and
a maximal amplitude of 79\,cm\,s$^{-1}$. The global results are in rather good agreement
with the recent work by Kjeldsen et al (\cite{kjel2}) who found large and small spacings
of 40.4 and 3.00\,$\mu$Hz. However, the comparison of the individual frequencies leads to
some discrepancies. Although the present data have a higher $S/N$, additional Doppler measurements 
using spectrographs like \textsc{Harps} (ESO) and forthcoming \textsc{Most} data with a clean window function
might help resolving these ambiguities.

We identified 22 mode frequencies which have been compared to theoretical models.
The combination of non--asteroseismic observations now available for \object{$\eta$ Boo} with the observed p--mode
frequencies listed in Table~\ref{tab:freq} leads to the following solution:
a model in the post-main-sequence phase of evolution, with a mass of $1.57 \pm 0.07$\,$M_{\odot}$, an age $t=2.67 \pm 0.10$\,Gyr and an initial metallicity $(Z/X)_{\mathrm{i}}=0.0391 \pm 0.0070$.
We also show that the mass of \object{$\eta$ Boo} is very sensitive to the choice of the observed metallicity and that 
its age depends on the inclusion of rotation and atomic diffusion. Indeed, non--rotating models
without overshooting predict a smaller age of $2.39 \pm 0.10$\,Gyr. 

\begin{acknowledgements}
We would like to thank J. Christensen--Dalsgaard for providing us with the Aarhus adiabatic pulsation code.
Part of this work was
supported financially by the Swiss National Science Foundation.
\end{acknowledgements}

\end{document}